\documentclass[12pt]{article}

\usepackage{scicite}

\usepackage{times}
\usepackage{graphicx,psfrag}
\usepackage[export]{adjustbox}
\usepackage{caption}
\usepackage{amsmath}

\newenvironment{sciabstract}{%
\begin{quote} \bf}
{\end{quote}}

\newcounter{lastnote}

\title{\large \bf  Anomalous refraction of acoustic guided waves in solids with geometrically tapered metasurfaces}

\author
{Hongfei Zhu$^{1}$ and Fabio Semperlotti$^{1,2^\ast}$ \\
\\
\normalsize{$^{1}$Department of Aerospace and Mechanical Engineering,}\\
\normalsize{University of Notre Dame, Notre Dame, IN 46556, USA}\\
\normalsize{$^{2}$ Ray W. Herrick Laboratories, School of Mechanical Engineering,}\\
\normalsize{Purdue University, West Lafayette, Indiana 47907, USA}\\
\normalsize{$^\ast$To whom correspondence should be addressed; E-mail:  fsemperl@purdue.edu}
}

\date{}

\begin{document}


\baselineskip24pt


\maketitle


\begin{sciabstract}
The concept of metasurface has recently opened new exciting directions to engineer the refraction properties in both optical and acoustic media. Metasurfaces are typically designed by assembling arrays of sub-wavelength anisotropic scatterers able to mold incoming wavefronts in rather unconventional ways. The concept of metasurface was pioneered in photonics and later extended to acoustics while its application to the propagation of elastic waves in solids is relatively unexplored. We investigate the design of acoustic metasurfaces to control elastic guided waves in thin-walled structural elements. These engineered discontinuities enable anomalous refraction of guided wave modes consistently with the generalized Snell's law. The metasurfaces are made out of locally-resonant torus-like tapers enabling accurate phase shift of the incoming wave which ultimately affects the refraction properties. We show that anomalous refraction can be achieved on transmitted antisymmetric modes ($A_0$) either when using a symmetric ($S_0$) or antisymmetric ($A_0$) incident wave, the former clearly involving mode conversion. The same metasurface design also allows achieving structure embedded planar focal lenses and phase masks for non-paraxial propagation.
\end{sciabstract}

\section*{Introduction}
The extensive development of engineered materials that has taken place during the last few decades provided new and unconventional ways to manipulate optical and acoustic waves. Beam forming and steering \cite{Amm1,beamform1,beamform2,beamform3}, lenses with super-resolution \cite{Superresolution1,Superresolution2,Superresolution3,Superresolution4}, and mechanical diodes\cite{MecDiode1,MecDiode2} are only a few of the many capabilities enabled by this new class of materials. As new applications for metamaterials are envisioned, a quest to reduce their size and develop subwavelength modulation devices begun. The concept of metasurface has recently emerged as a powerful approach to effectively manipulate wave-like fields while breaking the dependence on the propagation length. Metasurfaces are typically made of planar ultra-thin (that is subwavelength) gratings able to produce abrupt changes in the phase of the propagating waves. A common approach to the design of these elements exploits arrays of scatterers having both subwavelength separation and variable (either material or geometric) properties. The resulting interfaces exhibit either a spatially variable phase or impedance that virtually allows molding reflected and refracted wavefronts in any direction and shape. Compared to conventional optical and acoustic devices, such metasurfaces achieve wavefront control over distances much smaller than a wavelength.

Metasurfaces\cite{YuRev} were first studied in optics and implemented via optical antennas\cite{Yu,Ni} or microwave metamaterials\cite{Grady,Pfeiffer} in the context of light propagation across phase discontinuities. The Generalized Snell's Law (GSL) was introduced in order to predict the anomalous propagation across material interfaces characterized by a phase gradient\cite{Yu}. Inspired by this pioneering work, optical devices able to achieve unconventional wavefront manipulation capabilities have been theoretically and experimentally demonstrated. A few remarkable examples concern bending light in arbitrary shapes\cite{Ni,Huang}, conversion of propagating into surface modes\cite{Sun}, and the development of ultra-thin lenses\cite{Aieta,Kang}. Shortly after the introduction of metasurfaces in optics, the concept was adopted in acoustics with the intent of creating subwavelength acoustic devices. To-date, the most notable design of an acoustic metasurface is based on the labyrinthine, or space-coiling unit, which proved to be very effective for the control of pressure waves in gases. These labyrinthine structures have shown fascinating properties\cite{Liang,Li3,Liang2,Xie2,Tobias} including negative and zero effective refractive indices. Anomalous reflection and refraction have been both numerically\cite{Mei,Li1} and experimentally\cite{Li2,Xie,Tang,yifan,yuan} demonstrated with labyrinthine units based on phase discontinuities. For completeness, we note that acoustic metasurfaces based on impedance discontinuities (as opposed to phase discontinuities) have also been theoretically demonstrated\cite{Zhao1,Zhao2}.

Despite much progress in the development of these subwavelength designs in both optics and acoustics, the extension of this concept to the control of elastic waves in solids has not received much attention. Compared to acoustic waves in fluids and gases, elastic waves in solids exhibit a richer physics including the existence of different wave types and of mode conversion effects. The ability to extend the concept of metasurface to a solid could provide novel and important functionalities for structural acoustic waveguides while drastically expanding their range of application. These new capabilities could provide an unprecedented level of control on the path of energy propagation through a solid, ultimately enabling applications such as subwavelength ultrasonic lenses for imaging and biomedical applications (e.g. ultrasonic microscopy beyond the diffraction limit and ultrasonic surgery).

 In this article, we show both numerically and experimentally the possibility to successfully design acoustic metasurfaces fully embedded in structural waveguides and able to reliably achieve anomalous refraction. The metasurface design is based on the use of geometric tapers recently introduced by Zhu\cite{Zhu} in the context of thin-walled acoustic metamaterials. Previous results have shown that a periodic lattice of these geometric inhomogeneities can provide a high level of control of the spatial distribution of the propagation parameters due to the thickness-related dispersion behavior. At the same time, this design drastically reduces the fabrication complexity of traditional metamaterial systems based on a multi-material approach\cite{Ying,Lai,Liu}.

\section*{Acoustic metasurfaces based on geometric tapers}

\begin{figure}[h!]
\includegraphics[scale=0.8,center]{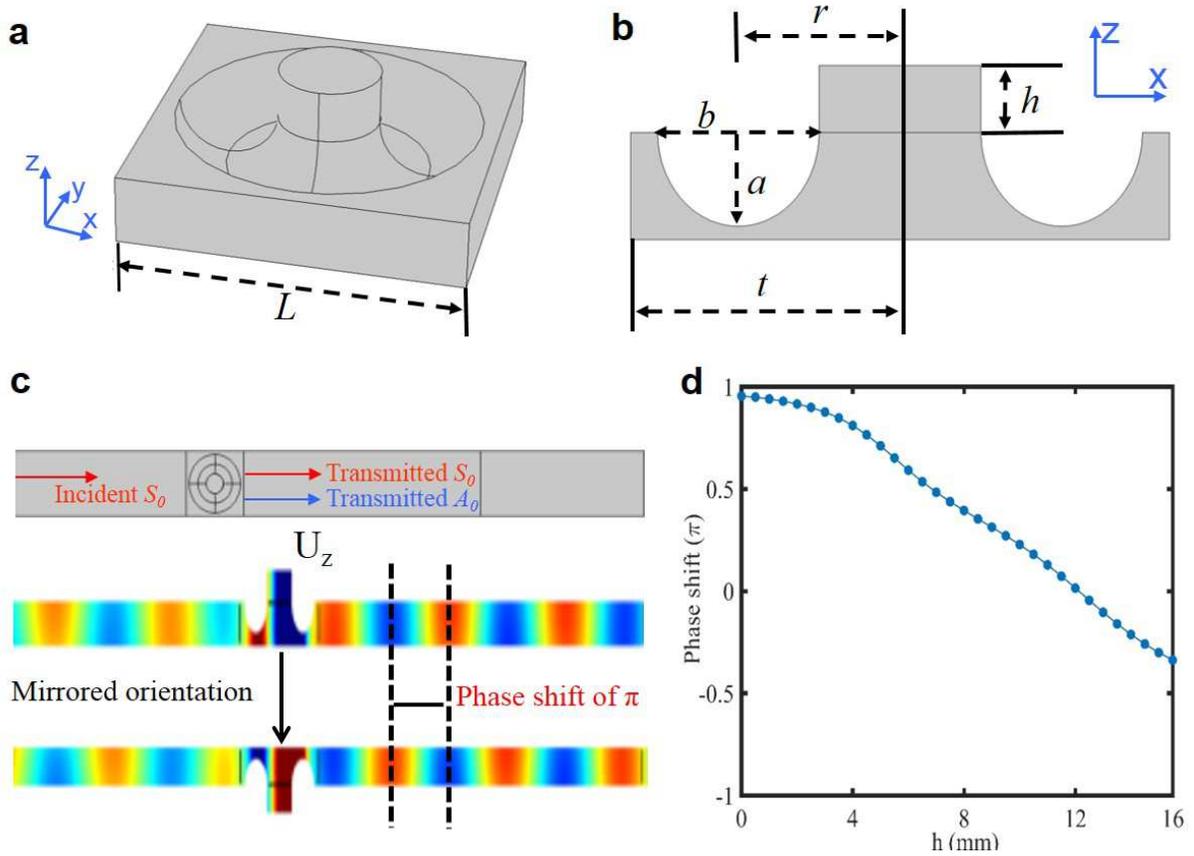}
   \caption{Schematic of the (a) elliptic torus-like tapered unit and of (b) its $xz$ plane cross-section showing the main geometric parameters. (c) Full field results illustrating the $\pi$ phase shift occurring on the transmitted (mode converted) $A_0$ mode generated by a normally incident $S_0$ mode when the same taper geometry is mirrored about the neutral plane. (d) The phase shift response of a single unit at $f=20.1$ kHz as a function of the thickness of the center mass (which results in a resonance frequency shift).} \label{Fig1}
\end{figure}

The proposed approach synthesizes the metasurface based on periodic arrays of elements whose individual unit consists in a geometrically inhomogeneous cell combined with a resonating core. In practice, each unit represents an anisotropic acoustic scatterer (Fig.\ref{Fig1}a). The scatterer consists of a torus-like taper which supports a center (resonating) mass. By tuning the parameters of the taper and of the mass in each individual unit, carefully engineered discontinuities can be designed and embedded into the supporting waveguide. In particular, the distribution of the units controls the spatial phase gradient along the metasurface and, ultimately, the refraction characteristics. In the following, we show by numerical simulations that anomalous refraction can be achieved on transmitted antisymmetric modes (e.g. $A_0$) either when using a symmetric (e.g. $S_0$) or an antisymmetric (e.g. $A_0$) incident wave; the former clearly involving mode conversion effects. This approach is then used to synthesize two types of compact acoustic devices such as (1) a flat ultra-thin lens, and (2) a phase-mask for non-paraxial propagation.

In order to predict the characteristics of reflected and refracted waves when in presence of inhomogeneous interfaces with phase gradients, we use the Generalized Snell's Law (GSL)\cite{Yu}. In the following, we concentrate on the refraction characteristics which are of greater interest for the design of practical devices. According to the GSL, the direction of anomalous refraction is related to the direction of the incident planar wavefront as follows:

\begin{align}
\frac{sin(\theta_t)}{\lambda_t}- \frac{sin(\theta_i)}{\lambda_i}=\frac{1}{2\pi}\frac{d\phi}{dx}\label{eqn1}
\end{align}

where $\theta$ indicates the angle of propagation, $\lambda$ the wavelength, $d\phi / dx$ the spatial phase gradient, and the subscripts $i$ and $t$ the incident and transmitted components of the wave. Equation (\ref{eqn1}) implies that the refracted beam can have an arbitrary direction, provided that a suitable constant phase-gradient ($d\phi / dx$) can be produced along the interface. In order to design metasurfaces that can effectively steer the transmitted guided wave, it is necessary to achieve a spatial gradient profile able to cover the entire $2\pi$ phase range. This condition translates into constraints on the frequency response of the different anisotropic scatterers. In particular, it requires that every scatterer achieves a full $2\pi$ phase change while simultaneously maximizing the amplitude of the transmitted wave (i.e. reducing the back-scattering). Mode conversion is one of the most characteristic traits of acoustic wave propagation in solids as compared to fluids or gases. In the present study, we show the possibility to achieve anomalous refraction for the $A_0$ anti-symmetric transmitted mode by illuminating the metasurface with either a $S_0$ or an $A_0$ incident mode. We will discuss the design for the $S_0$ excitation in detail while observing that the operating mechanism and the design strategy apply similarly to the $A_0$ mode.

\paragraph*{Design and frequency response of the fundamental unit cell.} The fundamental building block of the metasurface (Fig.\ref{Fig1}a) is a square unit (side length $L=4cm$) having an embedded elliptic torus-like taper and a (resonating) center mass whose value can be tuned by controlling its thickness. The $x-z$ cross-section of a typical unit is presented in Fig.\ref{Fig1}b, where $t$ is the position of the axis of symmetry of the center mass, $a$ and $b$ are the semi-major and semi-minor axis lengths of the elliptic torus, $r$ is the outer radius of the torus and $h$ is the thickness of the tunable center mass. In principle, there are many structural parameters that can be controlled to achieve specific amplitude and phase profiles. In this study, we mostly concentrated on the parameter $h$, which affects more directly the local resonance of the building block design, combined with different taper profiles. In order to evaluate the frequency response of the different units, we numerically calculated the transmitted $A_0$ mode resulting from an incident $S_0$ plane wave impinging normally on the metasurface. The phase and amplitude response were extracted from displacement data collected in the far-field. The frequency response characterization for each unit was performed on individual 3D strip-like models with a single embedded unit and periodic boundary conditions applied on the top and bottom boundaries, as shown in Fig.\ref{Fig1}c (top). For the sake of clarity and to facilitate the comparison between different units, we selected a specific frequency of actuation ($f=20.1$ kHz) which was used throughout the numerical simulations. This frequency corresponded to a wavelength of the $S_0$ mode $\lambda_{S_0} \approx $  27 cm $\sim 6.75 L$ and of the $A_0$ mode $\lambda_{A_0} \approx 5.9$ cm $\sim 1.5 L$ for a $8mm$ thick aluminum plate as considered in the numerical results.

In order to achieve phase shifts covering the entire $2\pi$ range when exciting the metasurface with a $S_0$ mode, we exploited mode conversion. The unit was excited by a $S_0$ plane wave at normal incidence traveling from left to right. Consider a pair of exactly identical tapered units, a $\pi$ phase shift difference in the transmitted and mode-converted $A_0$ wave can be achieved if the units orientation is mirrored with respect to the neutral plane (see Fig.\ref{Fig1}c). This mechanism can be easily understood by observing that the out-of-plane displacement field produced by the transmitted $A_0$ mode (anti-symmetric with respect to the neutral plane) must match the field produced by the incident $S_0$ wave in order to satisfy the continuity condition. the  direction. For one pair of units, the mirroring operation is equivalent to imposing the continuity conditions with an opposite sign, which leads in a straightforward way to a phase shift difference of $\pi$. These design considerations are confirmed by the finite element simulations presented in Fig.\ref{Fig1}c (bottom) that show the out-of-plane displacement field associated with the transmitted $A_0$ mode along the $xz$ cross-section of the strip model. The results of the two models, corresponding to the same taper geometry with mirror symmetry, clearly confirm the phase shift difference of $\pi$. We note that, the use of the mirrored geometry allows reducing the design complexity because the individual units are required to cover only a phase range of $\pi$, while the remaining ($\pi$) shift can be obtained by a mirroring operation.

In addition, the phase shift obtained by varying the thickness $h$ of the attached mass (up to twice the plate thickness) already produces phase shifts largely in excess of $\pi$ (Fig.\ref{Fig1}d). This means that the operating frequency of an individual unit can be further reduced therefore improving the subwavelength properties of the metasurface.

\begin{figure}[h!]
\includegraphics[scale=0.8,center]{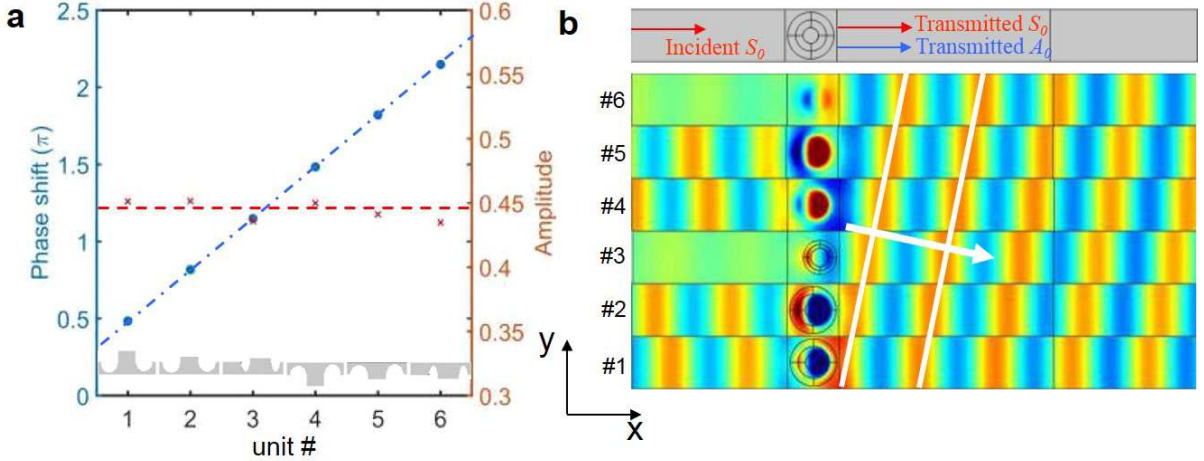}
   \caption{Phase, amplitude, and full field response produced by different units excited at a selected frequency $f = 20.1$ kHz. (a) Phase (blue dots) and amplitude (red dots) plots of the frequency response as a function of the unit geometry. The data are collected in the far-field of the transmitted $A_0$ (converted) mode when exciting the metasurface with a $S_0$ plane wave at normal incidence. The unit cross-sections in the inset of (a) correspond to the geometry modeled in (b). (b) Out-of-plane (z direction) displacement field produced by the different embedded units.} \label{Fig2}
\end{figure}

Following this design process, three pairs of basic units were selected. Figure \ref{Fig2}a shows the corresponding phase shift (blue dots) and amplitude (red dots) associated with the transmitted and mode-converted $A_0$ wave across each unit. Results indicate that the six discrete phase shifts cover the entire phase range in steps of approximately $\pi/3$ between adjacent units. The corresponding normalized amplitude for the transmitted converted $A_0$ mode is also sufficiently large with values fluctuating around 0.43, which is a normalized amplitude with respect to the incident $S_0$ wave. To provide further evidence of the phase-shift performance of the different units, the transmitted $A_0$ wave field is shown in Fig.\ref{Fig2}b. Each model provides the out-of-plane displacement pattern $U_z(x,y)$ at the neutral plane of the strip-like model. Results clearly show a cumulative phase shift covering the whole $2\pi$ range.

\section*{Numerical and Experimental Results}

In the following paragraphs, we explore the design of the embedded acoustic metasurface based on various combinations of the fundamental tapered units and able to achieve different forms of wave manipulation. In particular, we show the application of the metasurface to three phenomena: anomalous refraction, ultra-thin planar focusing lenses, and non-paraxial propagation. Note that, in the following numerical results the $S_0$ mode is always plotted in terms of the in-plane displacement field (x-component) at the neutral plane while the $A_0$ mode in terms of out-of-plane displacements (z-component).

\begin{figure}[h!]
\includegraphics[scale=0.8,center]{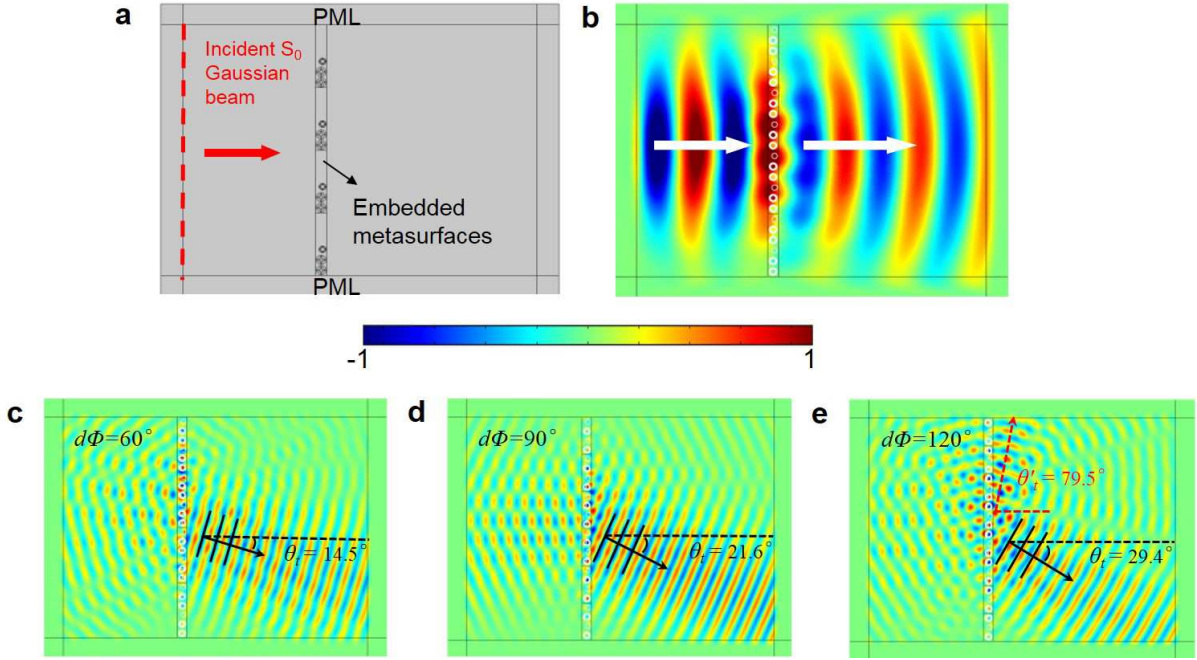}
   \caption{(a) Schematic of the thin plate with the embedded metasurface. A $S_0$ Gaussian beam with normal incidence was used as external excitation to excite the metasurface. (b) shows the ordinary refraction behavior of the transmitted $S_0$ mode; the visible reduction in the wave amplitude is due to the energy transferred into the anti-symmetric mode. (c)-(e) shows the transmitted (mode-converted) $A_0$ wave field for three metasurfaces having constant but different value of the spatial gradient. The resulting anomalous refraction is in excellent agreement with the prediction from the GSL indicated by the superimposed black arrows. } \label{Fig3}
\end{figure}

\paragraph*{Anomalous Refraction.} The fundamental characteristic of the metasurface design consists in the ability to achieve anomalous refraction. To illustrate this phenomenon we assembled a metasurface based on a one-dimensional periodic array of supercells. The individual supercell was built using different units properly selected to achieve a prescribed spatial phase gradient. The units were fully integrated in the host structure consisting in a $8mm$ thick aluminum plate, as shown in Fig.\ref{Fig3}a. The expected refraction angle $\theta_t$ can be predicted according to the GSL as $\theta_t=arcsin(\frac{\lambda_t sin(\theta_i)}{\lambda_i}+\frac{\lambda_t}{2\pi}\frac{d\phi}{dx})$.

To illustrate the occurrence of the anomalous refraction effect through the embedded metasurface we selected the case of normal incidence, i.e., $\theta_i=0$. The metasurface had a total length of $96$ cm equal to $24 L$, where $L$ is the in-plane dimension of the basic unit. The fundamental supercell at the basis of the metasurface resulted from the assembly of units with different geometry. The three different configurations provided a phase shift of $d\phi=\pi/3$, $d\phi=\pi/2$, $d\phi=2\pi/3$ along the interface. The metasurface was excited by a normally incident $S_0$ Gaussian beam. Figure \ref{Fig3}b shows the full field response of the $S_0$ mode which does not undergo refraction because the metasurface was designed to manipulate anti-symmetric modes. The visible reduction in the wave amplitude (around 40$\%$) across the metasurface is related to the portion of energy being converted into the $A_0$ mode. Figure \ref{Fig3}c-e show the transmitted $A_0$ mode for the different values of the phase gradient. The numerical results for every configuration show clear evidence of the anomalous refraction phenomenon. In addition, they are in excellent agreement with the analytical predictions provided by GSL according to which $\theta^{(60^\circ)}_t \approx 14.5^{\circ}$, $\theta_t^{(90^\circ)} \approx 21.6^{\circ}$ and $\theta^{(120^\circ)}_t \approx 29.4^{\circ}$. For better clarity, the analytical predictions are also superimposed on the full wave field results (black arrows in Fig. \ref{Fig3}c-e).
Only in the case of the strongest spatial gradient (Fig.\ref{Fig3}e), a visible portion of the incident beam was converted into a second refracted beam. This can be explained by the existence of a critical value for the phase discretization, that is the value of $d\phi$ between adjacent units. It can be shown that, beyond a certain value of $d\phi$ (for a given metasurface configuration in which the plate thickness $h$, the frequency of excitation, and the size of the fundamental array $dx$ are all kept constant; this condition is satisfied in our simulations) two possible propagating wave solutions can be supported by the same metasurface configuration. In fact, due to its wrap-around character, the phase shift of the metasurface can be written as $d\phi_n=d\phi\pm 2n\pi$, where $n$ is any integer. If the discretization is sufficiently fine (i.e. small $d\phi$), only $n=0$ can support a propagating wave solution satisfying $-1\le\frac{\lambda}{2\pi}\frac{d\phi_n}{dx}\le 1$ from GSL. However, if $d\phi$ is larger than a critical value, both the solution $n=0$ and $n=-1$ are possible. Following these considerations, the refraction angle of the second refracted beam can be determined by $\theta_{t}^{'}=arcsin{\frac{\lambda}{2\pi}\frac{-4/3\pi}{dx}}=-79.5^{\circ}$ which is in good agreement with the numerical simulation (see the dashed red arrow in Fig. \ref{Fig3}e). In our current configuration, the critical value for $d\phi$ can be calculated as $d\phi_c=2\pi-\frac{2\pi dx}{\lambda}=0.644\pi$. Aside from these additional considerations, results in Fig.\ref{Fig3} clearly illustrate that strong anomalous refraction is achievable on the $A_0$-converted Lamb modes with the proposed metasurface design.

\begin{figure}[h!]
\includegraphics[scale=0.8,center]{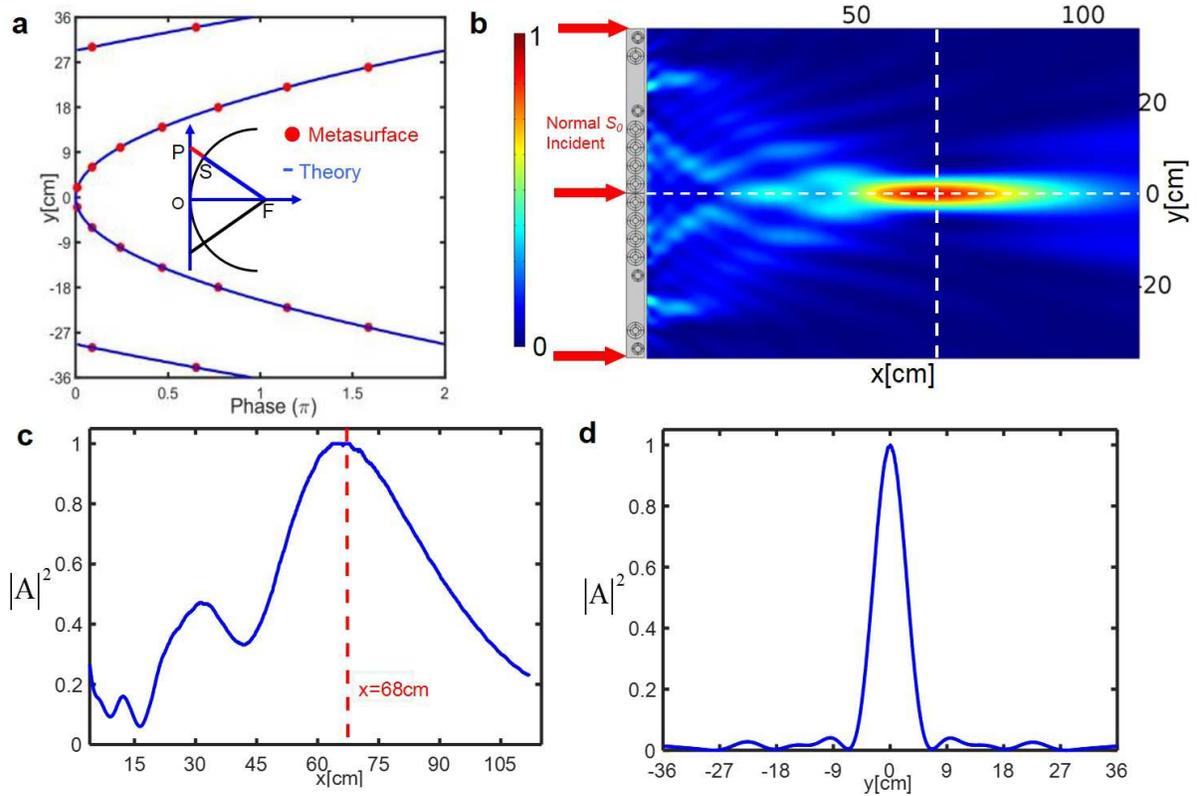}
   \caption{Design of a flat focal lens based on the embedded metasurface concept. (a) shows the continuous hyperbolic phase profile (blue curves) and the corresponding discrete phase profile (red dots) effectively implemented in the metasurface. The inset illustrates schematically how to synthesize the phase profile: the black circular line is the desired equal-phase contour for the transmitted $A_0$ mode showing that at point P the phase shift needs to be compensated proportionally to the distance $\overline{SP}$. (b) The intensity distribution $\vert A \vert ^2$ of the transmitted $A_0$ wave field after the flat focal lens.(c) Cross section of the intensity profile at $y=0$ cm. The peak appears at $x \approx 68$cm that is in very good agreement with the target value. (d) Transverse cross section of the intensity profile at $x=68$ cm.} \label{Fig4}

\end{figure}

\paragraph*{Flat Focal Lens.} Metasurfaces offer many possibilities to design planar and compact acoustic devices with diverse functionalities. Flat acoustic focal lenses are another example of devices achievable via geometrically tapered acoustic metasurfaces. To design such a flat lens, a hyperbolic phase profile must be programmed into the metasurface, as shown in Fig.\ref{Fig4}a. For a given focal length $f$, the phase shift $\phi(y)$ profile is provided by:
\begin{align} \label{flatlens}
\phi(y)=\vec{k}\cdot\overline{SP}=\frac{2\pi}{\lambda}(\sqrt{f^2+y^2}-f)
\end{align}

where $\vec{k}$ is the wave vector, $\overline{SP}$ is the distance to compensate for to achieve a hyperbolic profile, and $f$ is the focal length (that is the distance $\overline{OF}$).

To show the capabilities of the metasurface, we designed a lens having a focal length $f=70$ cm. The corresponding phase profile needed to achieve this performance was obtained according to Eqn.\ref{flatlens} and plotted in Fig.\ref{Fig4}a (blue curve). The metasurface relies on the use of discrete units to create an interface having a spatially varying phase profile, therefore the target hyperbolic phase profile can only be approximated by discrete values. For the generic $i$th element (which covers the section of the metasurface from $(i-1)L$ to $iL$), the phase shift is given by $\phi_i=\phi((i-1/2)L)$. The corresponding discrete phase profile (red dots) of the metasurface is plotted in Fig.\ref{Fig4}a and the lens can be readily constructed by selecting units that match the requested phase shift. Note that the lens was designed with the intent to use a normally incident $S_0$ excitation and to achieve a focal point in the transmitted $A_0$ (converted) mode. Figure \ref{Fig4}b shows the spatial distribution of the intensity of the transmitted wave field $\vert A \vert ^2$ after the metasurface. To quantify the performance of the acoustic lens, the parallel cross-section intensity distribution at $y=0$ cm (indicated by the white parallel dashed line in Fig.\ref{Fig4}b) is shown in Fig.\ref{Fig4}c. The peak occurs at $x\approx 68$cm that is within a 3 \% error from the target location ($f=70$cm). The intensity distribution along the transverse cross section at $x=68$ cm is also plotted in Fig.\ref{Fig4}d in order to show the very narrow focal point achievable with our metasurface concept.

\paragraph*{Design of a phase mask for non-paraxial propagation.} Geometrically tapered embedded metasurfaces are very versatile and can enable, among the possible applications, the design of \textit{phase masks} (essentially acoustic analog filters) to generate self accelerating acoustic beams that propagate along an arbitrary convex trajectory; a mechanism also known as non-paraxial propagation.

In particular, we design the metasurface to perform as a phase mask able to convert an incident wave into a non-paraxial beam. Based on the caustic theory\cite{Elad,Zhao3,Froehly}, the continuous spatial phase profile along the metasurface can be synthesized in order to achieve an arbitrary convex trajectory of the transmitted acoustic beam. Once the continuous phase profile is available, its discrete approximation can be easily obtained by selecting basic unit cells according to an analogous procedure to that described for the flat lens.

\begin{figure}[!ht]
\includegraphics[scale=0.8,center]{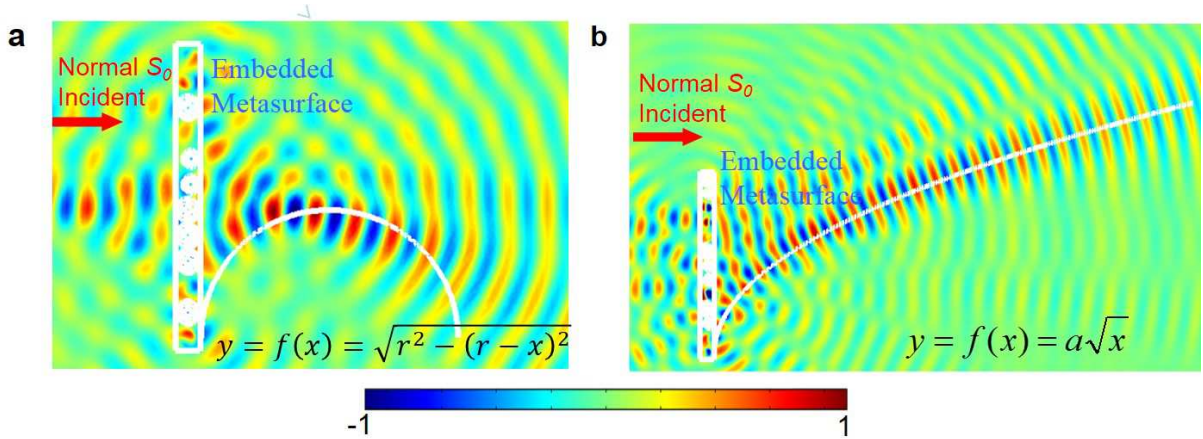}
   \caption{Non-paraxial self bending acoustic beam achievable by a metasurface phase mask. The metasurface, embedded into a $8mm$ thick plate, was designed to achieve a predefined phase profile yielding an arbitrary convex trajectory. A normally incident $S_0$ plane wave was used to excite the metasurface. Both designs can successfully achieve non-paraxial self-accelerating beams on the transmitted $A_0$ mode as shown in (a) for a half circular trajectory path and in (b) for a parabolic trajectory. The white solid lines in each plot represent the target trajectory.} \label{Fig5}
\end{figure}

The ability of the metasurface to generate non-paraxial beam propagation is numerically shown in the following by means of two different examples: (1) a half circular and (2) a parabolic trajectory. The half circular path with radius r and centered at $(r,0)$ is described by the equation $y=f(x)=\sqrt{r^2-(r-x)^2}$. The desired phase profile is $\phi(y_0)=-k_0*(y_0-2r*arctan(y_0/r))$. Concerning the parabolic trajectory $y=f(x)=a*\sqrt{x}$, the desired phase profile is $\phi(y_0)=-k_0*(-\ln(y_0+\sqrt{{y_0}^2+(a^2/4)^2}))*a^2/4$. For the numerical example, the parameters $r$ and $a$ controlling the curvatures of the paths were selected as $r=0.2 m$ and $a=\sqrt{0.32}$, in order to guarantee the smoothness of the discrete phase profile. The resulting metasurfaces were embedded in the host $8mm$ thick plate and excited by an incident $S_0$ plane wave. The numerical results are provided in Fig.\ref{Fig5}a and b in terms of full displacement fields associated with the transmitted (mode-converted ) $A_0$ wave. The superimposed white solid lines represent the target convex trajectories. In both cases, results show a very good agreement between the targeted and the calculated trajectories. Clearly the circular trajectory exhibits a larger error in the far quarter of the path due to the finite length of the metasurface.

\paragraph*{Achieving anomalous refraction on the transmitted $A_0$ waves without mode conversion.} The approach described above exploited the mode conversion mechanism, therefore the anomalously refracted wave had different polarization with respect to the incident one. The use of mode conversion allowed reducing the design complexity needed to achieve the $2\pi$ phase shift, however it was not strictly necessary to successfully build the metasurface. In the following, we show that embedded acoustic metasurfaces can also achieve anomalous refraction of transmitted $A_0$ waves when using an $A_0$ incident wave, that is without exploiting mode conversion.

\begin{figure}[h!]
\includegraphics[scale=0.6,center]{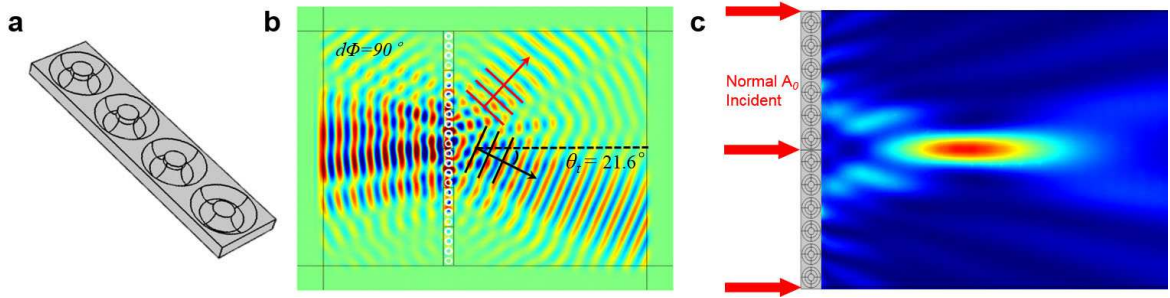}
   \caption{Metasurface design for an operating mode not involving mode-conversion. The metasurface is illuminated by an $A_0$ mode at normal incidence while the transmitted anomalously refracted beam maintains the same polarization. (a) detailed view of the geometry selected for the metasurface, (b) anomalous refraction of the transmitted $A_0$ mode, and (c) intensity distribution of a flat focal lens.} \label{Fig6}
\end{figure}

The design strategy follows the same approach illustrated above with the only exception that the incident wave is an $A_0$ mode and that the units phase shift must cover the whole $2\pi$ range, therefore the mirror symmetry cannot be leveraged. In this scenario, the design consists of an array of torus-like tapers with four different units that cover the $2\pi$ range phase shift in steps of $\pi/2$ (Fig.\ref{Fig6}a). The array is periodically repeated to form the metasurface and embedded into the host aluminum plate. To evaluate the performance, the metasurface was excited by an $A_0$ Gaussian beam with normal incidence. The high-amplitude beam in the area after the metasurface (Fig.\ref{Fig6}b) is well consistent with the anomalously refracted beam predicted by the GSL (represented by the black arrow superimposed on the full wave simulation results). A secondary low-amplitude beam indicated by the red arrow is also generated. This beam is likely due to a higher order of diffraction due to the supercell periodicity, explainable according to the diffraction grating theory\cite{Larouche}. In a similar way to what shown for the mode-conversion-based design, planar acoustic devices can also be developed. As an example, we show a flat focusing lens design whose field intensity distribution $\vert A \vert^2$ after the metasurface lens is presented in Fig.\ref{Fig6}c. Results clearly show the formation of a high intensity focal point when the metasurface is illuminated by an $A_0$ plane wave.

\paragraph*{Experimental Validation.} In order to validate the concept of geometrically tailored acoustic metasurface, we performed an experimental investigation targeted to the $A_0-A_0$ actuation mode. We selected this operating condition due to the simplicity of measuring the out-of-plane component of the vibration field and also because the $A_0$ is one of the most significant mode in thin-walled waveguides. The experimental set-up consists in a thin flat aluminum plate with a single embedded metasurface. In order to rescale the frequency of the $A_0$ mode in a range convenient for excitation and measurement, the plate thickness was reduced to $4mm$ and the basic unit size to $2cm\times2cm$. The target operating frequency for this design was set to $50.1$ kHz.

\begin{figure}[h!]
\includegraphics[scale=0.8,center]{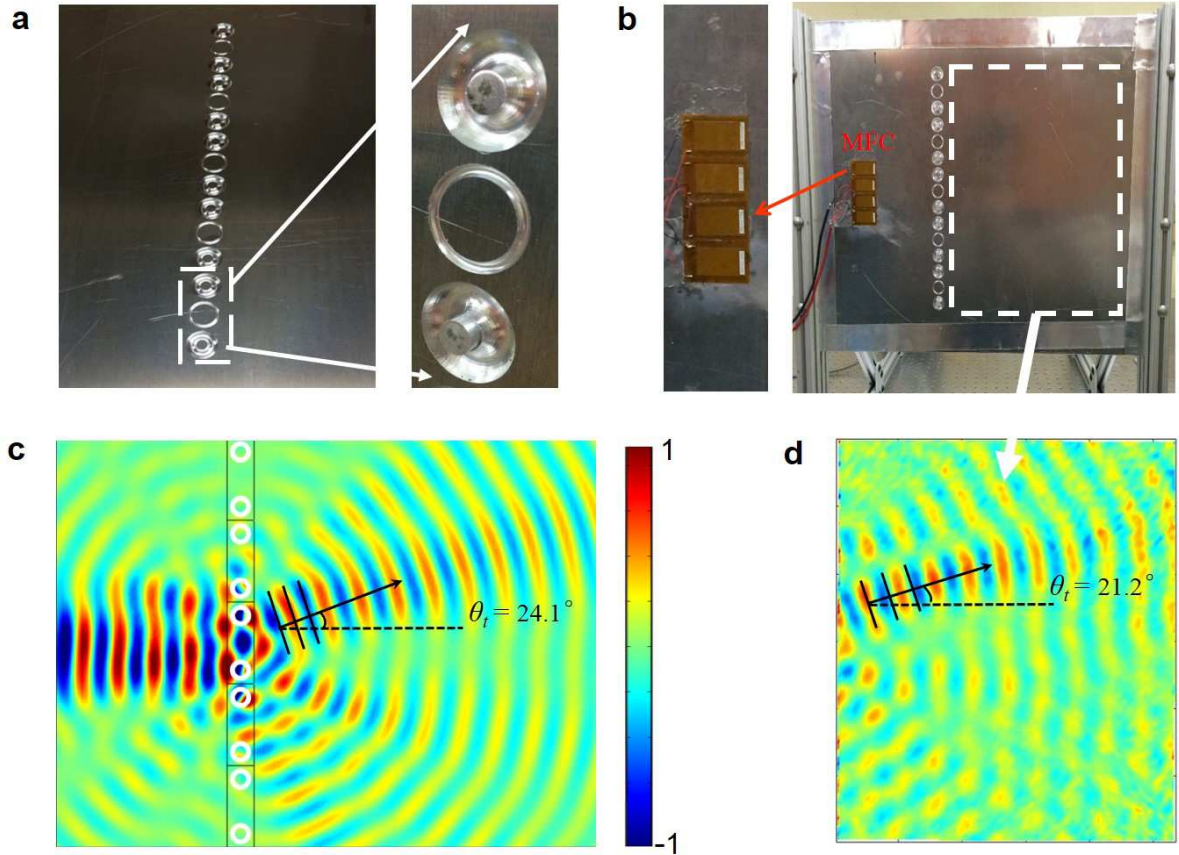}
   \caption{Experimental setup and results. (a) side view of the embedded metasurface on a $4mm$ thick aluminum plate. The zoom-in view shows the three unit supercell from which the periodic array is assembled. (b) Front-view of the setup showing the plate mounted in a supporting frame which provides clamped-free boundary conditions. The actuation was provided by four Micro Fiber Composite (MFC) patches glued on the surface of the panel and simultaneously actuated to generate an $A_0$ (quasi) plane wave. The white dashed box shows the area where measurements were collected via a Scanning Laser Vibrometer. (c) numerical results showing the predicted refracted beam for the selected configuration. (d) full field experimental measurement acquired on the test setup.} \label{Fig7}
\end{figure}

Numerical simulations equivalent to those illustrated above allowed selecting the metasurface configuration used for fabrication. The configuration consisted of three cells providing a phase increment of $\frac{2\pi}{3}$ between adjacent units. We highlight that this coarse discretization of the phase gradient was selected to simplify the experimental setup and the corresponding fabrication, and also to bring the resonance frequencies in a range convenient for measurement. Figure \ref{Fig7}a shows a view of the embedded metasurface assembled by repeating periodically the fundamental three-unit array as well as the detailed view of the array. The torus-like tapers were CNC machined into an initially flat plate while the center masses were fabricated separately and successively glued onto the taper.

The experimental sample was mounted in an aluminum frame to facilitate the measurements. The frame provided clamped-free boundary conditions on the edges of the plate. A 3M viscoelastic tape was used all around the panel to minimize the effect of boundary reflections. An array of four Micro Fiber Composite (MFC) patches was surface bonded on the plate and simultaneously actuated to generate an $A_0$ (quasi) planar incident wave. The MFC driving signal was a 50-count burst with a $50$kHz center frequency. The out-of-plane response of the plate was acquired by using a Polytec PSV-500 laser scanning vibrometer.

Numerical simulations (Fig.\ref{Fig7}c) were also conducted on the same configuration used for the experiments in order to obtain a prediction of the refracted beam. These results predicted an angle of refraction equal to 24.1$^{\circ}$. Figure \ref{Fig7}d shows the experimental measurement of the entire wave field in the area following the metasurface (white dashed box in Fig.\ref{Fig7}b) at a given time instant. The metasurface is able to generate an anomalous refracted beam at an angle 21.2$^{\circ}$. The numerical and experimental results (Fig.\ref{Fig7}c-d) are in very good agreement therefore proving the validity of the proposed design approach and the feasibility of using geometric tapers to create structure-embedded metasurfaces.

\section*{Conclusions}
We have presented and experimentally demonstrated a possible approach to the design of structure-embedded acoustic metasurfaces which can produce a variety of unconventional wave manipulation effects in structural waveguides. The geometrically tailored unit at the basis of the design proved to be very efficient, easy to tune, and able to yield compact metasurfaces for a variety of functionalities. The tapers are extremely easy to fabricate and eliminate completely the need for multi-material interfaces typical of the more traditional acoustic metamaterial design. This is a critical aspect to achieve scalability and to transfer metamaterial concepts to structural members with load-bearing capabilities.

Numerical and experimental results were in excellent agreement showing the robustness of the approach and the high level of performance achievable with tapered design. Acoustic wave control was shown for both \textit{same-mode} and \textit{mode-converted} transmitted waves. Applications of the metasurface to acoustic planar focal lenses and phase-masks for non-paraxial propagation were also successfully investigated by numerical simulations and showed an outstanding potential for wave control and device fabrication.

\bibliography{scibib}

\begin{thebibliography}{10}

\bibitem{Amm1}
Z.~Liu, {\it et~al.\/}, {\it Science\/} {\bf 289}, 1734 (2000).

\bibitem{beamform1}
S.~Fan, P.~R. Villeneuve, J.~D. Joannopoulos, H.~A. Haus, {\it Phys. Rev.
  Lett.\/} {\bf 80}, 960 (1998).

\bibitem{beamform2}
S.~Fan, {\it et~al.\/}, {\it Phys. Rev. B\/} {\bf 59}, 15882 (1999).

\bibitem{beamform3}
Y.~Pennec, {\it et~al.\/}, {\it Appl. Phys. Lett.\/} {\bf 87} (2005).

\bibitem{Superresolution1}
A.~Salandrino, N.~Engheta, {\it Phys. Rev. B\/} {\bf 74}, 075103 (2006).

\bibitem{Superresolution2}
Z.~Liu, H.~Lee, Y.~Xiong, C.~Sun, X.~Zhang, {\it Science\/} {\bf 315}, 1686
  (2007).

\bibitem{Superresolution3}
J.~de~Rosny, M.~Fink, {\it Phys. Rev. Lett.\/} {\bf 89}, 124301 (2002).

\bibitem{Superresolution4}
A.~Sukhovich, {\it et~al.\/}, {\it Phys. Rev. Lett.\/} {\bf 102}, 154301
  (2009).

\bibitem{MecDiode1}
F.~Li, P.~Anzel, J.~Yang, P.~G. Kevrekidis, C.~Daraio, {\it Nat. Comm.\/} {\bf
  5}, 5311 (2014).

\bibitem{MecDiode2}
N.~Boechler, G.~Theocharis, C.~Daraio, {\it Nat. Mat.\/} {\bf 10}, 665 (2011).

\bibitem{YuRev}
N.~Yu, F.~Capasso, {\it Nat. Mater.\/} {\bf 13}, 139  (2014).

\bibitem{Yu}
N.~Yu, {\it et~al.\/}, {\it Science\/} {\bf 334}, 333 (2011).

\bibitem{Ni}
X.~Ni, N.~K. Emani, A.~V. Kildishev, A.~Boltasseva, V.~M. Shalaev, {\it
  Science\/} {\bf 335}, 427 (2012).

\bibitem{Grady}
N.~K. Grady, {\it et~al.\/}, {\it Science\/} {\bf 340}, 1304 (2013).

\bibitem{Pfeiffer}
C.~Pfeiffer, A.~Grbic, {\it Phys. Rev. Lett.\/} {\bf 110}, 197401 (2013).

\bibitem{Huang}
L.~Huang, {\it et~al.\/}, {\it Nano. Lett.\/} {\bf 12}, 5750  (2012).

\bibitem{Sun}
S.~Sun, {\it et~al.\/}, {\it Nat. Mater.\/} {\bf 11}, 426  (2012).

\bibitem{Aieta}
F.~Aieta, {\it et~al.\/}, {\it Nano. Lett.\/} {\bf 12}, 4932 (2012).

\bibitem{Kang}
M.~Kang, T.~Feng, H.-T. Wang, J.~Li, {\it Opt. Express\/} {\bf 20}, 15882
  (2012).

\bibitem{Liang}
Z.~Liang, J.~Li, {\it Phys. Rev. Lett.\/} {\bf 108}, 114301 (2012).

\bibitem{Li3}
Y.~Li, {\it et~al.\/}, {\it Appl. Phys. Lett.\/} {\bf 101} (2012).

\bibitem{Liang2}
Z.~Liang, {\it et~al.\/}, {\it Sci. Rep.\/} {\bf 3} (2013).

\bibitem{Xie2}
Y.~Xie, B.-I. Popa, L.~Zigoneanu, S.~A. Cummer, {\it Phys. Rev. Lett.\/} {\bf
  110}, 175501 (2013).

\bibitem{Tobias}
T.~Frenzel, {\it et~al.\/}, {\it Appl. Phys. Lett.\/} {\bf 103} (2013).

\bibitem{Mei}
J.~Mei, Y.~Wu, {\it New. J. Phys.\/} {\bf 16}, 123007 (2014).

\bibitem{Li1}
Y.~Li, B.~Liang, Z.~Gu, X.~Zou, J.~Cheng, {\it Sci. Rep.\/} {\bf 3} (2013).

\bibitem{Li2}
Y.~Li, {\it et~al.\/}, {\it Phys. Rev. Applied\/} {\bf 2}, 064002 (2014).

\bibitem{Xie}
Y.~Xie, {\it et~al.\/}, {\it Nat. Comm.\/} {\bf 5} (2014).

\bibitem{Tang}
K.~Tang, {\it et~al.\/}, {\it Sci. Rep.\/} {\bf 4} (2014).

\bibitem{yifan}
Y.~Zhu, {\it et~al.\/}, {\it Sci. Rep.\/} {\bf 5} (2015).

\bibitem{yuan}
B.~Yuan, Y.~Cheng, X.~Liu, {\it Appl. Phys. Express.\/} {\bf 8} (2015).

\bibitem{Zhao1}
J.~Zhao, B.~Li, Z.~Chen, C.~W. Qiu, {\it Sci. Rep.\/} {\bf 3} (2013).

\bibitem{Zhao2}
J.~Zhao, B.~Li, Z.~N. Chen, C.-W. Qiu, {\it Appl. Phys. Lett.\/} {\bf 103}
  (2013).

\bibitem{Zhu}
H.~Zhu, F.~Semperlotti, {\it Phys. Rev. B\/} {\bf 91}, 104304 (2015).

\bibitem{Ying}
Y.~Wu, Y.~Lai, Z.-Q. Zhang, {\it Phys. Rev. Lett.\/} {\bf 107}, 105506 (2011).

\bibitem{Lai}
Y.~Lai, Y.~Wu, P.~Sheng, Z.-Q. Zhang, {\it Nat. Mat.\/} {\bf 10}, 620 (2011).

\bibitem{Liu}
X.~N. Liu, G.~K. Hu, G.~L. Huang, C.~T. Sun, {\it Appl. Phys. Lett.\/} {\bf 98}
  (2011).

\bibitem{Elad}
E.~Greenfield, M.~Segev, W.~Walasik, O.~Raz, {\it Phys. Rev. Lett.\/} {\bf
  106}, 213902 (2011).

\bibitem{Zhao3}
S.~Zhao, {\it et~al.\/}, {\it Sci. Rep.\/} {\bf 4} (2014).

\bibitem{Froehly}
L.~Froehly, {\it et~al.\/}, {\it Opt. Express\/} {\bf 19}, 16455 (2011).

\bibitem{Larouche}
S.~Larouche, D.~R. Smith, {\it Opt. Lett.\/} {\bf 37}, 2391 (2012).

\end{thebibliography}

\bibliographystyle{Science}

\end{document}